 \documentclass[final,3p,times]{elsarticle}
\usepackage{amssymb}
\usepackage{amsmath}
\usepackage{bm,bbm,hyperref,braket}
\usepackage{color,graphicx}
\usepackage{mathtools}

\biboptions{sort&compress}

\newcommand{\eq}[1]{\begin{align}#1\end{align}}

\newcommand{\mbf}{\mathbf}

\newcommand{\mr}{\mathrm}
\newcommand{\tx}{\text}

\journal{Chinese Journal of Physics}

\begin{document}

\begin{frontmatter}

\title{Testing trajectory-based determinism via probability distributions}

    \author[label1]{M. V. Scherer}
    \author[label1,label2]{A. D. Ribeiro}
    \author[label1]{R. M. Angelo}
    \affiliation[label1]{
        organization={Departamento de Física, Universidade Federal do Paraná},
%        addressline={Avenida Coronel Francisco H. dos Santos, 100}, 
        city={Curitiba},
        postcode={81531-990}, 
        state={Paraná},
        country={Brazil}}

\affiliation[label2]{
        organization={Department of Physics, University of Connecticut},
%             addressline={196 Auditorium Road},
             city={Storrs},
             postcode={06269},
             state={Connecticut},
            country={USA}}

%% Abstract
\begin{abstract}
%% Text of abstract
It is notorious that quantum mechanics cannot predict well-defined values for all physical quantities. Less well-known, however, is the fact that quantum mechanics is unable to furnish---without additional assumptions---probabilistic predictions even in emblematic scenarios such as the double-slit experiment. In contrast, trajectory-equipped theories naturally have more predictive power. This work formalizes the aforementioned assertions and illustrates them through three case studies: (i) free particle, (ii) free fall under a uniform gravitational field, and (iii) the double-slit experiment. Specifically, we introduce a prescription for constructing an arrival-time probability distribution within generic trajectory-equipped theories and then derive a conditional probability distribution that is unreachable by quantum mechanics. Our results can, in principle, be tested experimentally, thereby assessing the validity of trajectory-based determinism without the need for experiments involving the direct measurement of arrival time.
\end{abstract}

%%Graphical abstract
\begin{graphicalabstract}
\end{graphicalabstract}

%%Research highlights
\begin{highlights}
\item Standard Quantum Mechanics (SQM) does not assign a Hermitian operator for time, although there is a general consensus that this quantity can be measured.
\item Such an ambiguous SQM scenario also prevents us from deriving quantum distributions for time.
\item Trajectory-equipped theories (TET) naturally contemplate determinism, producing position-time and momentum-time correlations.
\item Using this framework, we deduce general probability distributions for time.
\item Employing TET, we suggest experimentally testing some predictions that are unreachable by SQM.
\end{highlights}

%% Keywords

\end{frontmatter}

%%%%%%%%%%%%%%%%%%%%%%%%%%%%%%%%%%%%%%%%%%%%%%%%%%%%%%%%%%%%
%%%%%%%%%%%%%%%%%%%%%%%%%%%%%%%%%%%%%%%%%%%%%%%%%%%%%%%%%%%%
\section{Introduction}
\label{intro}

\begin{figure}[t]
\centering
\includegraphics[width=11cm,angle=0]{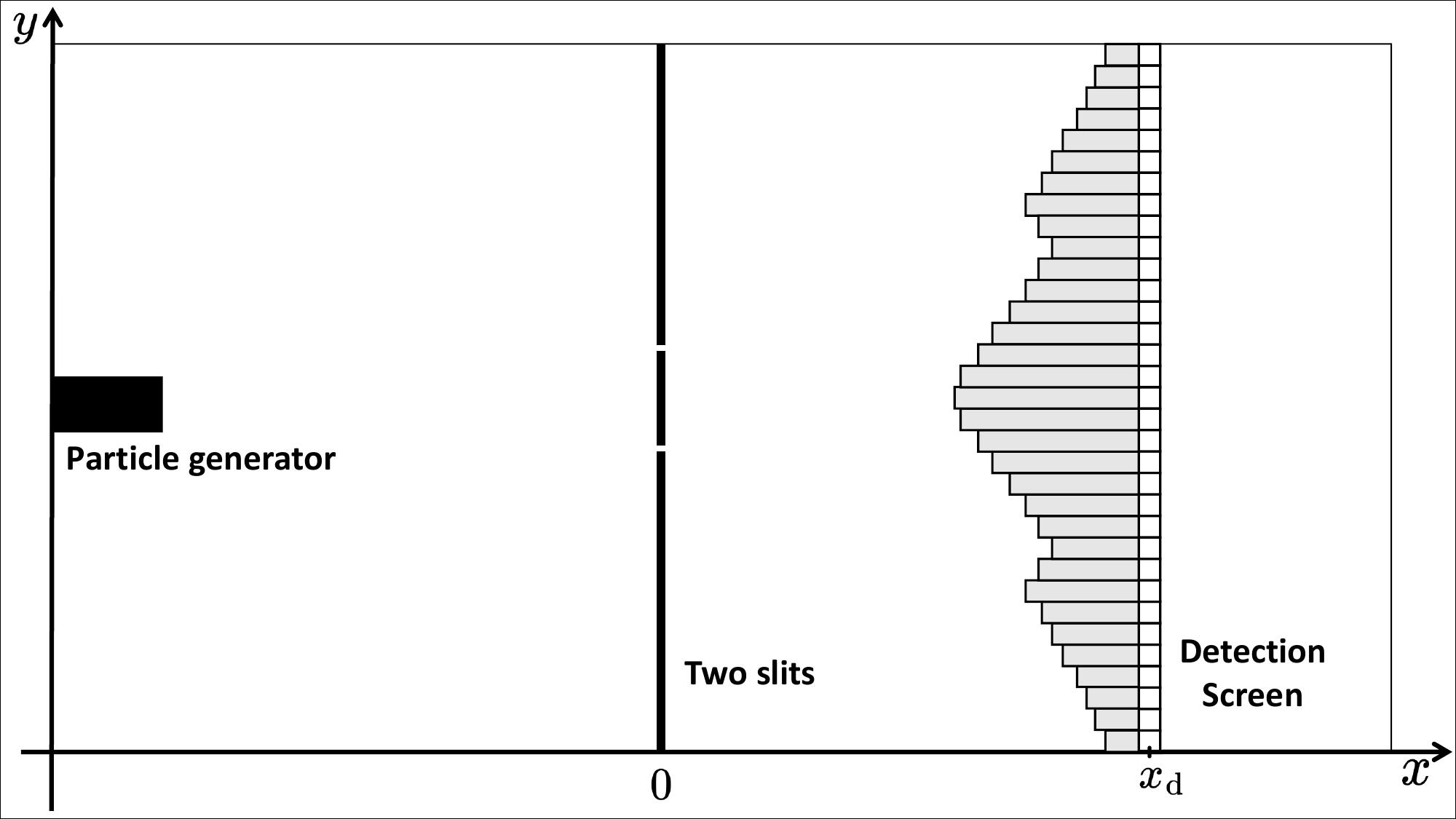}
\caption{Schematic representation of the double-slit experiment. The screen with the slits is located at $x = 0$, and the detection screen is placed at $x = x_{\text{d}}$, where each small square represents an individual detector. In this setup, particles are emitted one at a time, interact with the slits, and are eventually detected on the screen. By repeating the experiment many times, one can build a histogram of the conditional probability distribution ${\wp_{y|x}}(y|x_\tx{d})$, which quantifies the likelihood of detecting a particle at transverse position $y$, given the fixed screen location $x_\text{d}$. The gray bar plot shown in the figure schematically illustrates this distribution.}
\label{dupla_fenda}
\end{figure}

There are numerous physical and philosophical discussions related to quantum mechanics. However, one fact remains undeniable: the quantum formalism has consistently succeeded in describing the distribution of detector clicks in experiments. This success might lead some to believe that it provides objective answers (though not interpretative) for all observable phenomena. However, a critical examination of some traditional experiments reveals that this is not always true. For example, consider a double-slit experiment~\cite{Zeilinger, Gahler, Tonomura, Bach, Carnal, Kurtsiefer, Brand}, where the position $y$ of a particle is measured on a detection screen located at $x=x_\tx{d}$, with the sub-index ``$\tx{d}$'' referring to the detector. This setup is schematically shown in Fig.~\ref{dupla_fenda}. Typically, the particle's arrival time is neither controlled nor measured in such experiments. After many runs of the experiment, the observer constructs the experimental time-independent probability density ${\wp_{y|x}}^{\mkern-22mu\tx{exp}}(y|x_\tx{d})$, which describes the possible locations $y$ reachable by the particle, conditioned on the detection line $x=x_\tx{d}$. Now, how can one apply Born's rule from the theoretical joint probability density $|\psi(x,y,t)|^2$ to fit the experimental data?

%%%%%%%%%%%%%%%%%%
%%%%%%%%%%%%%%%%%%

A first move towards answering this question consists of recognizing that $|\psi(x,y,t)|^2$ is not a joint probability distribution for $x$, $y$, and $t$. Instead, it refers to the probability density of finding the system at the location $(x,y)$ {\it given} that time is guaranteed to be $t$. In this case, the notation $\wp_{xy|t}(x,y|t)$\footnote{To shorten the notation, whenever the sub-index of the symbol representing the probability density function repeats in its argument, we may omit the latter. For example, $\wp_{xy|t}(x,y|t)$ would be simply $\wp_{xy|t}$.}, emphasizing the conditioning on $t$, is preferred~\cite{Maccone, Dias_Parisio, Araújo, Arlans, Terno, Sombillo_Galapon,Rafsanjani_Bahrampour,Kellers}. Rather than just allegorical rephrasing, this serves to stress that the distribution has to be built for a single well-defined value of time. Next, we can condition on $x=x_\tx{d}$:
\eq{\label{p(y|x,t)}
\wp_{y|xt}(y|x_\tx{d},t)=\frac{\wp_{xy|t}(x_\tx{d},y|t)}{\wp_{x|t}(x_\tx{d}|t)}=
\frac{|\psi(x_\tx{d},y,t)|^2}{\int\! dy\,|\psi(x_\tx{d},y,t)|^2},
}
where $\wp_{y|xt}(y|x_\tx{d},t)$ is the probability density of finding the system at the transversal position $y$, conditioned to the time instant $t$ and $x=x_\text{d}$, while $\wp_{x|t}(x_\tx{d}|t)$ refers to the probability density of measuring it at $x=x_\text{d}$, in a given time $t$. Unless one post-selects the experimental data to a particular value of $t$, which actually is rarely done in practice \cite{Sinha}, we still do not have a formal comparison with ${\wp_{y|x}}^{\mkern-22mu\tx{exp}}(y|x_\tx{d})$. Of course, our first impulse would be to assign some mean value for time in Eq.~\eqref{p(y|x,t)} or even to integrate over some time window, but these would be extra assumptions. This underscores that experimental results cannot always be straightforwardly confronted with quantum mechanics predictions. 

Experiments like the ones quoted above are usually carried out in the far-field regime, so semiclassical analysis generally applies \cite{Vona,Das,Shucker,Wolf}. Nevertheless, recent advancements in ultra-fast detector technology \cite{Korzh, Steinhauer, Azzouz, Rosticher} have made it possible to explore regimes where semiclassics is insufficient to describe the system's behavior \cite{Vona, Delgado}. Therefore, it is mandatory to develop a strategy to deal with time. A viable route to this end would be to search for a theoretical probability distribution for the arrival time, $\wp_{t|x}(t|x_\tx{d})$, and then use probability theory to obtain the desired distribution,
\eq{\label{DP_yx}
\wp_{y|x}(y|x_\tx{d}) = \int\! dt\, \wp_{y|xt}(y|x_\tx{d},t)\, \wp_{t|x}(t|x_\tx{d}).
}
It turns out, however, that the canon of quantum mechanics (QM) alone (i.e., without supplementary hypotheses\footnote{By ``supplementary hypotheses'' we mean all assumptions beyond the axioms of the theory, such as those involving semiclassics and time quantization (for some of these approaches, see Ref.~\cite{Roncallo2023}).}) lacks a well-defined and widely accepted prescription for the construction of $\wp_{t|x}(t|x_\tx{d})$. Pauli's theorem~\cite{teorema_Pauli} establishes the impossibility of defining a Hermitian operator corresponding to the arrival time within the framework of QM. As a consequence, predictions cannot be formulated using standard projective measurements. Within the broader scope of generalized measurements, typically implemented via the formalism of positive operator-valued measures (POVMs), the situation appears more promising, though it remains controversial. In Ref.~\cite{Anastopoulos2006}, for instance, the authors adopted the consistent histories framework to derive an arrival-time POVM which, although it retrieves the one proposed by Kijowski~\cite{Kijowski} for free particles, suffers from the quantum Zeno effect and strongly depends on an arbitrary temporal resolution of the measuring device. On the other hand, some approaches construct arrival-time POVMs by \textit{ad hoc} quantization of free-particle trajectories (see Ref.~\cite{Muga_Egusquiza} and references therein), while more recent results indicate that such POVMs may not even exist in certain systems involving particles with spin~\cite{Goldstein2024-1}.

As a first contribution of this article, we demonstrate that {\em general} trajectory-equipped theories can address this issue, allowing for a probability distribution for arrival time, as they naturally accommodate the concept of determinism, correlating space and time variables at a functional level. After presenting a general formalism, we then provide some illustrations in terms of, e.g., classical statistical mechanics and Bohmian mechanics (BM) \cite{Bohm_I,Bohm_II}---the latter being a well-known model that reinstates determinism while upholding probabilistic predictions consistent with QM. Although different from ours in some important aspects, approaches are already known that use Bohmian trajectories to construct time probability distributions~\cite{Leavens_II,Leavens,Muga_Leavens,Leavens_McKinnon,Leavens_McKinnon_II,Leavens_III,Leavens_IV,Alonso,Das_Durr,Holland}. In fact, several trajectory-based models exist~\cite{Nelson,Hall,McLafferty} that allow access to the probability distribution of time~\cite{Nitta, Kazemi, Rafsanjani}. Other approaches to time in QM are worth citing. While the Kijowski method~\cite{Kijowski,Muga_Egusquiza} employs semiclassical approximations to derive temporal predictions, the Page-Wooters mechanism~\cite{Page_Wootters} considers time as an inaccessible coordinate that arises from correlations between subsystems of the global state. Similar approaches conceiving time as an emergent property of physical systems can be found in Refs.~\cite{Briggs_Rost, Dias, Rost_Gemsheim}. The spacetime-symmetric extension of QM~\cite{Dias_Parisio} treats time on par with space, resulting in a Schr\"odinger equation for time. This approach has found application to the problem of the tunneling time~\cite{Ximenes}. Finally, there are phenomenological modeling approximations of the detection process that yield probability distributions of time, such as the absorbing boundary rule~\cite{Werner,Tumulka_I,Tumulka_II,Dubey}, path-integral with absorbing boundary~\cite{Marchewka_I,Marchewka_II,Marchewka_III}, and others~\cite{Juric_I,Damborenea,Muga_Brouard,Halliwell}. 

Given the frequent use of Bohmian trajectories in the present work, it is important to mention a recent debate within a scenario involving both spatial and spin degrees of freedom in relativistic BM, which has been established regarding the measurability of arrival time and the relevance of the detection system in the very characterization of this time~\cite{Goldstein2024-1,Goldstein2024-2,Das2024,Drezet}. Our work avoids those issues by adhering to the following premises. First, we consider only the spatial degrees of freedom of a single particle in the non-relativistic regime. Second, we adopt the perspective that detectors, being rigidly fixed in the reference frame, have classical space-time degrees of freedom (we always know where they are and with what velocity because they are part of our reference frame)~\cite{Dieguez2018}. In this regard, principles of relationality forbid one from describing the dynamics of those degrees of freedom. This places serious restrictions on credible unitary descriptions of detectors (which, in a way, touches on Bohr's philosophy regarding these objects). Most importantly, recent evidence suggests that no microscopic account of the detector (and no additional free parameters) is necessary for BM to provide a good fit to experimental data~\cite{Das2022}. Here, we assume that the same holds true for all trajectory-equipped theories, at least at an approximative level.

Relying on the above assumptions, we then develop a formalism based on the hypothesis of trajectory-based determinism, aiming to derive spatial and temporal probability distributions. More specifically, this formalism enables the computation of probability distributions $\wp_{x|t}$ and $\wp_{t|x}$, and finally, the one in Eq.~\eqref{DP_yx}, which turns out to be the central figure of merit for experimental tests. We then employ BM to give a concrete example of the formalism usage and compare our results with those of other approaches~\cite{Leavens_II,Leavens, Muga_Leavens,Bracken,Hofmann}. Besides this introduction, the text is organized as follows. In Sec. \ref{sec:tet}, we introduce the basis for our model, which involves two postulates and the development of conditional probability distributions, one of them for the arrival time. In Sec.~\ref{sec:Studies}, we apply our formalism to some emblematic physical problems and compare our results with other approaches (when available). We close the article in Sec.~\ref{sec:conclusion} with our final remarks.

%%%%%%%%%%%%%%%%%%%%%%%%%%%%%%%%%%%%%%%%%%
\section{Trajectory-equipped theory (TET)}
\label{sec:tet}

In this section, we present our formalism, treating separately two types of deterministic trajectories, namely, those defined in terms of a velocity field and those in phase space. To simplify the discussion, we restrict ourselves to the one-dimensional case. The key idea consists of inverting the trajectory to express time as a function of the initial condition, whose statistics are governed by a well-defined probability distribution. Several works have also utilized this approach~\cite{Leavens_II,Leavens,Muga_Leavens,Leavens_McKinnon,Leavens_McKinnon_II,Leavens_III,Leavens_IV,Alonso,Das_Durr,Holland}, but the focus was on Bohmian trajectories, typically without recurrence points, and applied to the tunneling time problem. As we detail below, our formalism is more general, as it applies to generic trajectories, including those defined in phase space and with an arbitrary number of recurrences.

%%%%%%%%%%%%%%%%%%%%%%%%%%%%%%%%%%%%%%%%%%%%%%%%%%%%%%%%%%
\subsection{Trajectories associated with a velocity field}
\label{sub:TwFoVA}

In this scenario, trajectories emerge from a first-order differential equation $\dot{q}=f(q,t)$, where $f$ is some real function representing the velocity field. Setting $q_0$, the coordinate of the particle at $t=0$, the coordinate $q$ of the particle at the instant $t$ is uniquely defined. As is the case in BM, $q_0$ is the only boundary condition needed, and the mechanical momentum of a particle of mass $m$ is just given by $p=m f(q,t)$. Now, as far as the triple $\{q,q_0,t\}$ is concerned, here is the catch: just as the specification of $q_0$ and $t$ implies $q$, the specification of $q$ and $q_0$ (respectively, $q$ and $t$) should imply $t$ (respectively, $q_0$). In addition, due to the functional structure connecting the triple, one variable inherits the uncertainties of the others, provided physical principles are not violated. We rigorously formalize these ideas as follows:
\begin{itemize}
\item {\it Determinism hypothesis}.---There exists a smooth trajectory $q = Q(q_0,t)$ connecting an initial coordinate $q_0$ to a point $q$ over a time lapse $t$. For bounded dynamics, $q$ is expected to be a recurrence point, reachable at several instants $t_j\in\tau_n\equiv\{t_1, t_2, \dots, t_n\}$. Additionally, in typical well-behaved dynamics, $Q$ is invertible (at least in parts). Therefore, there must be functions $Q_0$ and $\{T_{\mkern-4mu j}\}_{j=1}^n$ allowing us to express determinism in terms of the relations
\eq{q = Q(q_0,t),\quad
    q_0 = Q_0(q,t),\quad
    t_j = T_{\mkern-4mu j}(q, q_0),
}
with $t\in\tau_n$. In this trajectory-equipped theory (TET), the probability distributions for finding the system at the final location $q$, given the initial location $q_0$ and time $t$, and for finding it at time $t$, given the initial and final locations $q_0$ and $q$, are
\begin{subequations}\label{hip_1}
\eq{
& \wp_{q|q_0t}(q|q_0,t) = \delta\big(q-Q(q_0,t)\big), \label{hip_1_1} \\
& \wp_{t|qq_0}(t|q,q_0) = \frac{1}{n} \sum_{j=1}^{n}  \delta\big(t-T_{\mkern-4mu j}(q,q_0)\big),\label{hip_1_2}
}
\end{subequations}
respectively, where $\delta$ denotes the Dirac delta function. As expected, here we have non-negative distributions satisfying $\int\!dq\,\wp_{q|q_0t}(q|q_0,t_j) = \int_0^{\infty}\!dt\,\wp_{t|qq_0}(t|q,q_0) = 1.$ By allowing for the possibility of recurrence points in the trajectory, we naturally incorporate the phenomenon of backflow into our theory.

\item {\it Preparation independence hypothesis}.--- The probability distribution for $q_0$ derives from some experimental procedure that is in no way physically conditioned to $q$ and $t>0$, hence
\eq{\label{hip_2}
 \wp_{q_0|qt}(q_0|q,t)= \wp_{q_0}(q_0) \equiv \rho_0(q_0).
}
The preparation $\rho_0(q_0)$ should be specified in each statistical theory. In QM, $\rho_0(q_0)=\langle q_0|\hat{\rho}_0|q_0\rangle$, with $\hat{\rho}_0$ the density operator at $t=0$.
\end{itemize}

In what follows, we will build upon the previously mentioned assumptions, using the definition of marginal probability,
\eq{\label{def_prob_marginal}
\wp_{v|r}(v|r) = \int\!du\,\wp_{v|ru}(v | r,u)\,\wp_{u|r}(u|r) ,
}
along with the property of the Dirac delta function,
\eq{\label{propri_delta}
\delta\left( g(u) \right) = \sum_{i=1}^N  \frac{\delta(u-u_i)}{|\partial_u g(u)|_{u=u_i}},
}
where $\partial_ug\equiv\partial g/\partial u$ and $g(u_i)=0$ for $1\leq i \leq N$. Plugging the distributions \eqref{hip_1_1} and \eqref{hip_2} into Eq.~\eqref{def_prob_marginal}, and solving the integral over $q_0$ with the property \eqref{propri_delta}, we obtain the probability density of finding the system at the final location $q$ conditioned on time $t$
\eq{\label{mean_1} 
\wp_{q|t}(q|t) =\left[\frac{ \rho_0(q_0) }{\left| \partial_{q_0} Q(q_0,t) \right|}\right]_{q_0=Q_0(q,t)}. 
}
The modulus operation appearing in the denominator guarantees that $\wp_{q|t} \geq 0$. The probability distribution for arrival time is constructed similarly. Using the relations \eqref{hip_1_2}-\eqref{propri_delta}, we end up with the probability density of finding the system at the time $t$ given the final location $q$
\eq{\label{mean_2}
\wp_{t|q}(t|q) =  \frac{1}{n}\sum_{j=1}^{n} \left[\frac{\rho_0\big( q_0\big)}{ \left| \partial_{q_0} T_{\mkern-4mu j}(q, q_0) \right|}\right]_{q_0=Q_{0j}(q,t)},
}
where the function $Q_{0j}(q,t)$ is achieved from manipulating the equation $t = T_{\mkern-4mu j}(q, q_0)$ so that $\smash{q_0=Q_{0j}(q,t)}$. As should be clear by now, our formalism [summarized by Eqs.~\eqref{mean_1} and \eqref{mean_2}] requires the existence of these analytic functions for the trajectories. This is a limiting factor in some situations, such as, for instance, within the Bohmian framework in the presence of wave-function nodes, where invertible functions are not available, or chaotic trajectories, which would demand a proper numerical treatment.

For future reference, we now indicate how the probability distributions of arrival time and positions can be associated with each other, under specific conditions. Starting with $\wp_{q|q_0t}$, we can use the property \eqref{propri_delta} to write
\eq{\label{eq_Q}
\wp_{q|q_0t}  = \delta \big(q- Q(q_0,t) \big) = \sum_{j=1}^{n} \frac{ \delta \big( t - T_{\mkern-4mu j}(q, q_0)  \big)}{ v_j(q,q_0) }, 
}
where we have introduced $v_j(q,q_0) \equiv |\partial_{t} Q(q_0,t) |_{t= T_{\mkern-4mu j}(q, q_0)}\geq 0$. Multiplying Eq.~\eqref{eq_Q} by $\rho_0(q_0)$ and integrating over $q_0$ gives 
\eq{\label{eq_geral_t}
\wp_{q|t}(q|t) =  \sum_{j=1}^{n} \left[\frac{\rho_0( q_0)}{ v_j(q,q_0) \left| \partial_{q_0} T_{\mkern-4mu j}(q, q_0) \right|} \right]_{q_0=Q_{0j}(q,t)}. 
}
In the specific case where $v_j(q,q_0) \equiv v(q,t)$, independently of~$j$, we find, from Eq.~\eqref{mean_2}, 
\eq{ \wp_{t|q}(t|q) = \frac{v(q,t)\,\wp_{q|t}(q|t)}{n}.
}
Finally, specializing to BM, $v(q,t)$ is readily identified with the velocity field, and $|J(q,t)|\equiv v(q,t) \, \wp_{q|t}(q|t)$ with the absolute value of the probability current density. In addition, when $q$ is not a recurrence point $(n=1)$, the last equation becomes
\eq{\label{densidade_corrente}
 {\wp_{t|q}}^{\scriptscriptstyle \mkern-25mu\tx{L}}\,\,(t|q)=|J(q,t)|,
}
which is the same expression derived by Leavens around two decades ago~\cite{Leavens_II,Leavens,Muga_Leavens} (notice the symbol ``L'' to refer to this author). In fact, it is crucial to realize that our approach generalizes the existing one in two aspects: it applies to all trajectory-equipped theories, not only to BM, and it naturally accommodates the backflow effect in terms of trajectories and velocity field.

%%%%%%%%%%%%%%%%%%%%%%%%%%%%%%%%%%%%%
\subsection{Phase space trajectories}
\label{sub:TwoFoVA}

We now consider trajectories defined in the phase space $qp$ through real functions $q=\mathcal{Q}(\mbf{r}_0,t)$ and $p=\mathcal{P}(\mbf{r}_0,t)$, constrained by the initial condition $\mbf{r}_0=(q_0,p_0)$ at $t=0$. Importantly, since $q$ and $p$ are treated as independent canonical variables, the uncertainties associated with $q_0$ and $p_0$ can similarly be independent. The initial probability distribution is denoted as $\rho_0(\mbf{r}_0)$ or, equivalently, $\rho_0\big(q_0,p_0\big)$. For simplicity, in what follows, we assume that there are no recurrences (i.e., the trajectory passes through the point $\mbf{r}=(q, p)$ only once), so the inversion of the trajectories yields only the functions $t =\mathcal{T}_{\mkern-4mu q}(q, \mbf{r}_0)$ and $t = \mathcal{T}_{\mkern-5mu p}(p,\mbf{r}_0)$. Again, we propose an approach relying on two fundamental assumptions.
\begin{itemize}
\item {\it Determinism hypothesis}.---There exists a trajectory in phase space defined by smooth functions $q=\mathcal{Q}(\mbf{r}_0,t)$ and $p=\mathcal{P}(\mbf{r}_0,t)$, connecting an initial point $\mbf{r}_0=(q_0,p_0)$ to a final point $\mbf{r}=(q, p)$ over a time lapse $t$. Assuming that the involved functions are invertible and that there are no recurring points, the determinism hypothesis can be expressed by the relations
\eq{\label{const_q}
\begin{array}{lll}
q =\mathcal{Q}(\mbf{r}_0,t), & &q_0 =\mathcal{Q}_{0q}(q,p_0,t),\\ [4pt]
p_0 =\mathcal{P}_{0q}(q,q_0,t), & &t = \mathcal{T}_{\mkern-4mu q}(q, \mbf{r}_0),
\end{array}
}
and
\eq{\label{const_p}
\begin{array}{lll}
p =\mathcal{P}(\mbf{r}_0,t),&  & q_0 = \mathcal{Q}_{0p}(p,p_0,t), \\ [4pt] 
p_0 =\mathcal{P}_{0p}(p,q_0,t), & & t = \mathcal{T}_{\mkern-5mu p} (p,\mbf{r}_0).
\end{array}
}
Note that the symbol $\mathcal{Q}_{0q}$ (respectively, $\mathcal{P}_{0q}$) refers to a function originating from the inversion of $q =\mathcal{Q}(\mbf{r}_0,t)$, with output $q_0$ (respectively, $p_0$). Similar reasoning has motivated us to name the other functions appearing in Eqs.~\eqref{const_q} and~\eqref{const_p}. In addition, note that the pairs $\big\{\mathcal{T}_{\mkern-4mu q}, \mathcal{T}_{\mkern-5mu p}\big\}$, $\big\{\mathcal{Q}_{0q}, \mathcal{Q}_{0p}\big\}$, and $\big\{\mathcal{P}_{0q}, \mathcal{P}_{0p}\big\}$ consist of distinct functions. Based on these relations, we propose the following conditional probabilities:
\begin{subequations}\label{determinism_hypothesis_2}
\eq{
& \wp_{qp|\mbf{r}_0t} = \delta\big(q-\mathcal{Q}(\mbf{r}_0,t)\big)\, 
\delta\big(p-\mathcal{P}(\mbf{r}_0,t)\big),\label{qp} \\
& \wp_{qt|p\mbf{r}_0} = \delta\big(q-\mathcal{Q}(\mbf{r}_0,t)\big)\, 
\delta\big(t-\mathcal{T}_{\mkern-5mu p}(p,\mbf{r}_0)\big),\label{qt}\\
& \wp_{pt|q\mbf{r}_0} = \delta\big(p-\mathcal{P}(\mbf{r}_0,t)\big)\, 
\delta\big(t-\mathcal{T}_{\mkern-4mu q}(q,\mbf{r}_0)\big).\label{pt}
}
\end{subequations}
The approach is such that starting with two postulated functions $\mathcal{Q}$ and $\mathcal{P}$ involving the variables $\{q,p,q_0,p_0,t\}$, one can construct a joint probability distribution for two of them out of the specification of the others. Note that while the distribution~\eqref{qp} admits the interpretation $\wp_{qp|\mbf{r}_0t}=\wp_{q|\mbf{r}_0t} \,\wp_{p|\mbf{r}_0t}$, meaning that $q$ and $p$ are uncorrelated variables both conditioned to $\{\mbf{r}_0,t\}$, the same does not hold for the ones involving time. Equation~\eqref{qt}, for instance, can be interpreted as $\wp_{qt|p\mbf{r}_0}=\wp_{q|\mbf{r}_0t} \,\wp_{t|p\mbf{r}_0}$, with the probability for one of the variables, $\wp_{q|\mbf{r}_0t}$, depending explicitly on the other. All this is consistent with the premise that $q$ and $p$ are independent canonical variables, while $q$ and $t$ are not.

\item {\it Preparation independence hypothesis}.---The joint probability for $\mbf{r}_0$ derives from some experimental procedure that is in no way physically  conditioned to $\mbf{r}$ and $t>0$, hence
\eq{\label{rho0(x0)}
 \wp_{\mbf{r}_0|\mbf{r}\mkern1mu t}(\mbf{r}_0|\mbf{r},t)= \wp_{\mbf{r}_0}(\mbf{r}_0) \equiv \rho_0(\mbf{r}_0).
}
The preparation $\rho_0(\mbf{r}_0)$ should be specified in each statistical theory. In QM, one possibility is to use the Husimi distribution~\cite{Husimi}, $\langle \alpha|\hat{\rho}_0|\alpha\rangle/\pi$, where $|\alpha\rangle$ denotes a coherent state with complex label $\alpha=\alpha(\mbf{r}_0)$ parametrized in terms of phase-space coordinates.
\end{itemize}

Utilizing the same procedure as before, we integrate the product of distributions \eqref{qp} and \eqref{rho0(x0)} over $\mbf{r}_0$ to obtain, with the aid of property~\eqref{propri_delta},
\eq{\label{mean_class_1}
\wp_{qp|t}(q,p|t) = \left[\frac{ \rho_0\big(\mathcal{Q}_{0q}(q,p_0,t),p_0\big) }
{ \mathcal{D}_{q_0}^\mathcal{Q}(q,p_0,t) ~
\mathcal{D}_{p_0}^\mathcal{P}(q,p_0,t)}\right]_{p_0 = \mathcal{P}_{\mkern-6mu\mathcal{Q}}(q,p,t)},
}
where 
\eq{
\begin{aligned}
\mathcal{D}_{q_0}^\mathcal{Q}(q,p_0,t) &\equiv 
\left| \partial_{q_0}\mathcal{Q}\big(q_0,p_0,t\big)\right|_{q_0=\mathcal{Q}_{0q}(q,p_0,t)}, \\
\mathcal{D}_{p_0}^\mathcal{P}(q,p_0,t) &\equiv
\left| \partial_{p_0}\mathcal{P}\big(\mathcal{Q}_{0q}(q,p_0,t),p_0,t\big)\right|, 
\end{aligned}
}
and $p_0=\mathcal{P}_{\mkern-6mu\mathcal{Q}}(q,p,t)$ is found by isolating $p_0$ in the equation $p=\mathcal{P}\big( \mathcal{Q}_{0q}(q,p_0,t),p_0,t\big)$. Via similar procedures, from Eq.~\eqref{qt}, we obtain
\eq{\label{mean_class_2}
\wp_{qt|p}(q,t|p) = \left[\frac{ \rho_0\big(\mathcal{Q}_{0q}(q,p_0,t),p_0\big) }
{\mathcal{D}_{q_0}^\mathcal{Q}(q,p_0,t) ~
\mathcal{D}_{p_0}^{\mathcal{T}}(q,p,p_0,t)}\right]_{p_0 = \mathcal{P}_{\mathcal{T}}(q,p,t)},
}
where
\eq{
\mathcal{D}_{p_0}^\mathcal{T}(q,p,p_0,t) \equiv
\left| \partial_{p_0}\mathcal{T}_{\mkern-5mu p}\big(p,\mathcal{Q}_{0q}(q,p_0,t),p_0\big)\right| 
}
and $p_0 = \mathcal{P}_{\mathcal{T}}(q,p,t)$ is found by isolating $p_0$ in the equation $t=\mathcal{T}_{\mkern-5mu p}\big(p,\mathcal{Q}_{0q}(q,p_0,t),p_0\big)$, and, from Eq.~\eqref{pt}, 
\eq{\label{mean_class_3}
\wp_{pt|q}(p,t|q) = \left[\frac{ \rho_0\big(\mathcal{Q}_{0q}(q,p_0,t),p_0\big) }
{ \mathcal{D}_{q_0}^{\mathcal{T}}(q,p_0,t) ~
\mathcal{D}_{p_0}^\mathcal{P}(q,p_0,t)}\right]_{p_0 = \mathcal{P}_{\mkern-6mu\mathcal{Q}}(q,p,t)},
}
where function $\mathcal{P}_{\mkern-6mu\mathcal{Q}}$  is same one that appears in Eq.~\eqref{mean_class_1} and
\eq{
\mathcal{D}_{q_0}^\mathcal{T}(q,p_0,t) \equiv
\left| \partial_{q_0}\mathcal{T}_{\mkern-4mu q}\big(q,q_0,p_0\big)\right|_{q_0=\mathcal{Q}_{0q}(q,p_0,t)}. 
}

The equations \eqref{mean_class_1}, \eqref{mean_class_2}, and \eqref{mean_class_3} provide instructions for deriving probability distributions for a trajectory in phase space. To find them, specifically in the form of $\wp_{q|t}$ and $\wp_{t|q}$, marginalization should be performed with respect to the momentum~$p$.

%%%%%%%%%%%%%%%%%%%%%%
\section{Case studies}
\label{sec:Studies}

In this section, we apply our formalism to simple physical systems and, when pertinent, compare our results with those of other approaches. However, we point out that QM does not provide an equivalent formulation for the expressions achieved here. Therefore, in this sense, there is no theoretical framework strong enough to be considered as a benchmark, reinforcing that the final verdict about the validity of our method should necessarily involve experimental results. In the absence of a formula to assume as correct, even the derivation process of the analytical expressions performed in Sect.~\ref{sub:TwFoVA} and ~\ref{sub:TwoFoVA} can be questioned. For this reason, as a self-consistency test, we also developed a numerical-statistical method to check their correctness. Essentially, this method counts the number of trajectories connecting the initial and final constraints relevant to each application. For instance, suppose that we are working in the context of trajectories generated from a velocity field, described in Sect~\ref{sub:TwFoVA}. Then, we assume knowing an initial probability density of initial positions $\wp_{q_0|t}(q_0|0)$, and our goal is calculating the time distribution $\wp_{t|q}(t|q_{\mathrm{d}})$, conditioned to the detector position $q_{\mathrm{d}}$. To accomplish this task without using Eq.~\eqref{mean_2}, but still considering the {\em determinism} and {\em preparation independence} hypotheses, we can apply the following reasoning. First, we construct an ensemble of $N_t$ initial positions $q_0$, respecting the distribution $\wp_{q_0|t}(q_0|0)$, namely, the number of points inside the interval $[q_0,q_o+dq_0]$ must be proportional to $\wp_{q_0|t}(q_0|0)$. Thus, we time-evolve each trajectory from this ensemble, and when it arrives at $q={q}_{\mathrm{d}}$, we record the value of $t$ for this event. At last, we simply count the number of trajectories that arrive at ${q}_{\mathrm{d}}$ inside the arrival-time interval $[t,t+dt]$ to get the distribution $\wp_{t|q}(t|{q}_{\mathrm{d}})$. Notice that this method is totally independent of Eq.~\eqref{mean_2}, but we expect that both results agree at least in the limit $N_t\to\infty$. Again, we point out that this {\em counting method} is useful only to check the consistency of the analytical expressions derived for such probability density functions.

%%%%%%%%%%%%%%%%%%%%%%%%%%
\subsection{Free particle}
\label{sub:FP}

We start considering the one-dimensional dynamics of a free particle with mass $m$, prepared in a Gaussian state written as $\psi(q,t=0) = \sqrt{G_q(\bar{q}_0,\sigma_0)}\,\mbox{e}^{i\bar{p}_0 \left( q-\Bar{q}_0 \right)/\hbar}$, where 
\eq{\label{gaussiana} G_u(\bar{u},\Delta)\coloneqq \frac{1}{\sqrt{ 2\pi \Delta^2}} \exp{\left[ - \frac{(u-\bar{u})^2}{2\Delta^2} \right]}.
}
The parameter $\bar{q}_0$ ($\bar{p}_0$) is the mean position (momentum) and $\sigma_0$ is the spatial uncertainty. For the role of TET, we choose two participants: classical statistical mechanics (or Liouvillian mechanics) and BM. For the classical model, the phase-space trajectory is given by 
\eq{\label{traj_class}
q &= \mathcal{Q}(\mbf{r}_0,t) = q_0 + \frac{p_0t}{m}, \nonumber \\
p &=  \mathcal{P}(\mbf{r}_0,t) = p_0, \nonumber
}
while $\rho_0(\mbf{r}_0)=G_{q_0}\left(\bar{q}_0,\sigma_0\right)\,G_{p_0}\left(\bar{p}_0,\frac{\hbar}{2\sigma_0}\right)$ stands for the phase-space initial distribution. The joint probability distribution for the position and momentum $(q,p)$, given the time $t$, is then obtained using Eq.~\eqref{mean_class_1}. The result reads 
\eq{
{\wp_{qp|t}}^{\scriptscriptstyle \mkern-40mu\tx{C}\mkern+28mu}(q,p|t)= 
\rho_0\bigg(q-{\textstyle\frac{pt}{m}},p\bigg)= 
G_q\bigg(\bar{q}_t, \sigma_0\bigg)\, G_p \bigg(\bar{p}_0, {\textstyle\frac{\hbar}{2\sigma_0}}\bigg),  \nonumber
}
where $\bar{q}_t \equiv \bar{q}_0 + \frac{pt}{m}$ and ``$\mr{C}$'' stands for ``classical". Marginalizing over $p$ gives the probability distribution for finding the final position $q$ given the arrival time $t$
\eq{
{\wp_{q|t}}^{\scriptscriptstyle \mkern-25mu\tx{C}\mkern+13mu}(q|t)= 
\int\! dp \, {\wp_{qp|t}}^{\scriptscriptstyle \mkern-40mu\tx{C}\mkern+28mu}(q,p|t)  
= G_q \left(\bar{q}_0 + \frac{\bar{p}_0t}{m}, \sigma_t\right),
}
where
\eq{\label{sigma_omega}
\sigma_t = \sigma_0 \sqrt{1+(\omega t)^2}, \qquad \omega=\hbar/(2m\sigma_0^2).
}
As expected, these results plainly match the ones obtained from the Liouvillian formalism, where ${\wp_{qp|t}}^{\scriptscriptstyle \mkern-40mu\tx{C}\mkern+28mu}$ emerges from the application of the inverse map of the trajectory to the initial probability distribution~\cite{greiner}. In addition, ${\wp_{q|t}}^{\scriptscriptstyle \mkern-25mu\tx{C}\mkern+13mu}=|\psi(q,t)|^2$, meaning that the classical distribution mirrors the quantum one, as already known for the free particle system. 

\begin{figure}[t]
\centering
\includegraphics[width=11cm,angle=0]{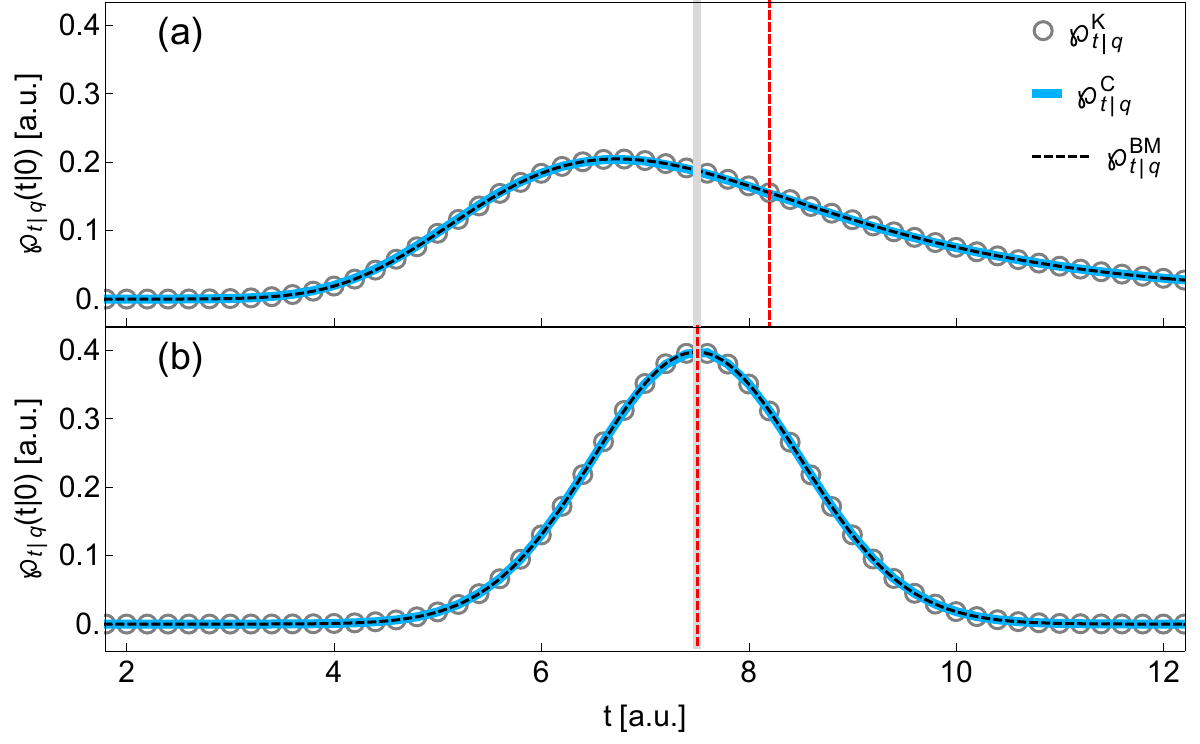}
\caption{Probability distributions ${\wp_{t|q}}^{\scriptscriptstyle \mkern-25mu\tx{C}\mkern+13mu}(t|0)$ (cyan solid line), ${\wp_{t|q}}^{\scriptscriptstyle \mkern-25mu\tx{BM}\mkern+1mu}(t|0)$ (black dashed line), and ${\wp_{t|q}}^{\scriptscriptstyle \mkern-25mu\tx{K}\mkern+13mu}(t|0)$ (gray circles), for the arrival time of a free particle detected at the position $q=0$, according to two trajectory-equipped theories, the classical one (C) and Bohmian mechanics (BM), and the Kijowski method (K). Panels (a) and (b) depict distributions for two parameter sets, namely, ${ \{\sigma_0=2,\bar{q}_0=-15,\bar{p}_0 =1, m=0.5,\hbar=1\} }$ and ${ \{ \sigma_0=2,\bar{q}_0=-15,\bar{p}_0 =20,m=10, \hbar=1 \} }$, respectively. All physical quantities and distributions are in arbitrary units. The gray vertical lines denote the arrival time $-m\bar{q}_0/\bar{p}_0$ for the center of the classical distribution, while the red dashed lines represent the average times, which converge to the same value for all distributions. A complete agreement is observed among the models. All the results presented in this graph come from well-defined continuous functions.}
\label{f1}
\end{figure}

To compute the arrival-time probability distribution, we start from Eq.~\eqref{mean_class_3} to obtain
\eq{\label{pCt-free}
{\wp_{pt|q}}^{\scriptscriptstyle \mkern-40mu\tx{C}\mkern+28mu}(p,t|q)= 
\frac{|p|}{m}\,
G_q\bigg(\bar{q}_t, \sigma_0\bigg)\, G_p \bigg(\bar{p}_0, {\textstyle\frac{\hbar}{2\sigma_0}}\bigg).
}
Now, to get ${\wp_{t|q}}^{\scriptscriptstyle \mkern-25mu\tx{C}\mkern+13mu}$, one needs to integrate this result over $p$. Due to its complexity, the analytical result will be omitted, but it is numerically illustrated in Fig.~\ref{f1}, constrained to the detection position $q=0$. These graphs will be discussed later when confronted with the next approaches.

%%%%%%%%%%%%%%%%%%
%%%%%%%%%%%%%%%%%%

%%%%%%%%%%%%%%%%%%%
\begin{figure}[tb]
\centerline{
\includegraphics[width=11cm,angle=0]{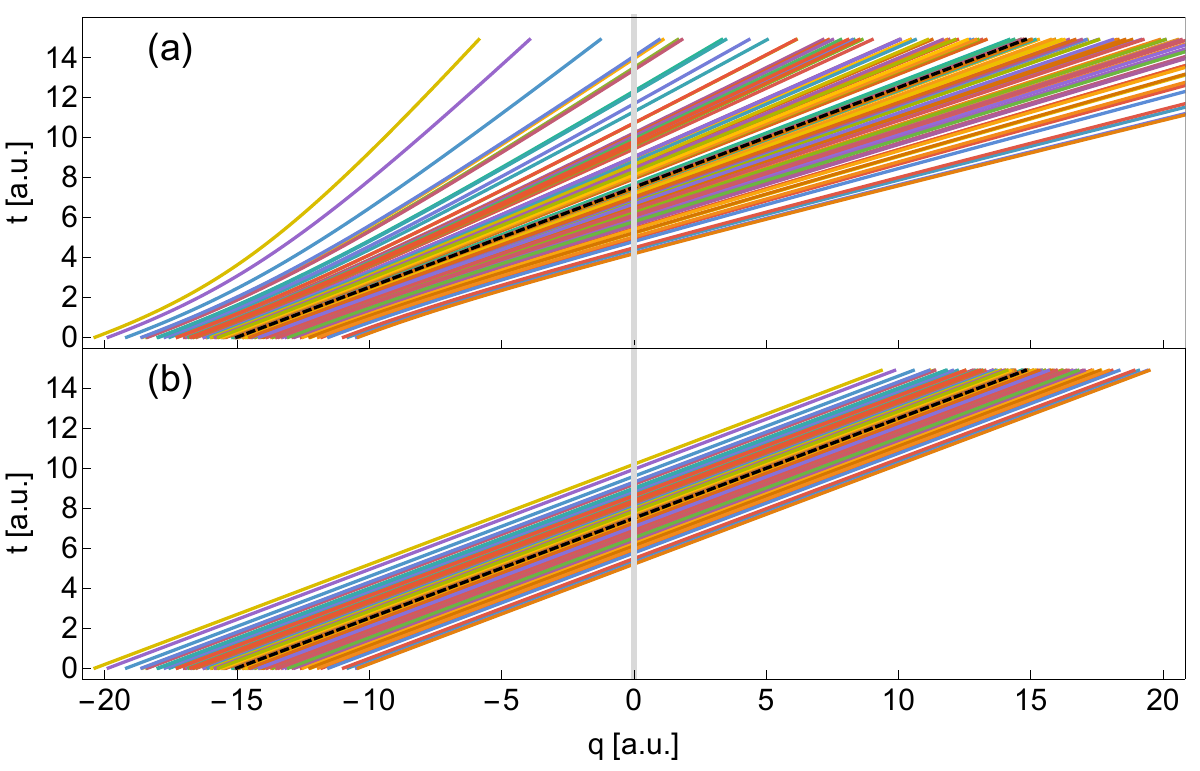}}
\caption{Bohmian trajectories for a free particle. They were obtained from $N_t=100$ initial positions $q_0$, chosen according to the distribution $|\psi(q_0,0)|^2$. Both panels utilize the same ensemble of initial conditions. The trajectory starting from ${q_0=\bar{q}_0}$ is represented by the black dashed curve. Panels (a) and (b) depict the distributions for two sets of parameters: ${ \{\sigma_0=2,\bar{q}_0=-15,\bar{p}_0 =1, m=0.5,\hbar=1\} }$ and ${ \{ \sigma_0=2,\bar{q}_0=-15,\bar{p}_0 =20,m=10, \hbar=1 \} }$, respectively. The vertical gray line at $q=0$ indicates the position where the detector is located. All physical quantities are given in arbitrary units.}
\label{f}
\end{figure}
%%%%%%%%%%%%%%%%%%%

As a second TET, we consider BM. According to Bohmian theory, the trajectory for the free particle is given by~\cite{Leavens_II}
\eq{\label{BT-free}
Q(q_0,t)=  \bar{q}_0 + \frac{\bar{p}_0t}{m} + \sqrt{1+(\omega t)^2}\left( q_0 - \bar{q}_0 \right).
}
To illustrate this kind of dynamics, in Fig.~\ref{f}, an ensemble of Bohmian trajectories~\eqref{BT-free} is showcased for two distinct sets of parameters. The choice of $q_0$ for each trajectory was taken considering the initial distribution $|\psi(q_0,0)|^2$, which implies that the density of the initial points is proportional to this function. Now, with $Q(q_0,t)$ and its inverse $Q_0(q,t)$, we come to Eq.~\eqref{mean_1} to show that ${\wp_{q|t}}^{\scriptscriptstyle \mkern-25mu\tx{BM}\mkern+1mu}={\wp_{q|t}}^{\scriptscriptstyle \mkern-25mu\tx{C}\mkern+13mu}=|\psi(q,t)|^2$. Through Eq. \eqref{mean_2}, we derive the arrival-time distribution given the final position $q$:
\eq{\label{pMBt-free}
{\wp_{t|q}}^{\scriptscriptstyle \mkern-25mu\tx{BM}\mkern+1mu}(t|q) =
 \left| \frac{\bar{p}_0}{m} + \frac{(\omega t)^2}{1+(\omega t)^2} \left( \frac{q-\bar{q}_0}{t} - \frac{\bar{p}_0}{m}\right)\right|\, |\psi(q,t)|^2,
 }
where the term inside the modulus is the velocity field $v(q,t)$. Analyzing Eq.~\eqref{pMBt-free}, we identify that ${\wp_{t|q}}^{\scriptscriptstyle \mkern-25mu\tx{BM}\mkern+1mu}(t|q)=|J(q,t)|$, where $J$ is the probability current density. This is expected because the Bohmian trajectory for the free particle has no recurrence points. Thus, the distribution converges to the Leavens expression~\cite{Leavens_II,Leavens,Muga_Leavens,Leavens_McKinnon,Leavens_McKinnon_II,Leavens_III,Leavens_IV,Alonso,Das_Durr,Dumont}, as shown in Eq.~\eqref{densidade_corrente}.

We also bring to the discussion the distribution deriving from the Kijowski method~\cite{Kijowski,Muga_Egusquiza}, which relies on the hypothesis of a generalized measurement operator for arrival time. Specifically for the arrival location $q=0$ and Gaussian state, the resulting arrival-time probability distribution reads~\cite{Muga_Egusquiza}
\eq{\label{pKt-free}
\begin{aligned}
{\wp_{t|q}}^{\scriptscriptstyle \mkern-25mu\tx{K}\mkern+13mu}(t|0)&= 
\frac{\sigma_t}{m\pi \hbar^2  \sqrt{2\pi}} \sum_{\lambda=\pm 1} \bigg|\int \! dp  \,
\bigg[\Theta(\lambda p) \sqrt{|p|}
%  \\ &\quad\times
 ~\mbox{e}^{i(p-\bar{p}_0)\bar{q}_0/\hbar}\,  \mbox{e}^{-\sigma_t^2 (p-\bar{p}_0)^2/\hbar^2}\, 
\mbox{e}^{ip^2t/(2m\hbar)} \bigg] \bigg|^2,
\end{aligned}
}
where $\Theta$ is the step function. In our simulations, the integral was computed numerically using the Monte Carlo method.

In Fig.~\ref{f1}, we depict the distributions ${\wp_{t|q}}^{\scriptscriptstyle \mkern-25mu\tx{C}\mkern+13mu}(t|0)$, ${\wp_{t|q}}^{\scriptscriptstyle \mkern-25mu\tx{BM}\mkern+1mu}(t|0)={\wp_{t|q}}^{\scriptscriptstyle \mkern-25mu\tx{L}}\,\,(t|0)$, and ${\wp_{t|q}}^{\scriptscriptstyle \mkern-25mu\tx{K}\mkern+13mu}(t|0)$ for two sets of parameters (the same as those of Fig.~\ref{f}). The first remark is that all theories yield precisely the same distribution. To a certain extent, this is not surprising, as the model under inspection is the simplest possible test: the dynamics is linear, and the preparation is Gaussian. On the other hand, there is subtler information coming from the results. In the two panels considered, the mean velocities are identical, suggesting that the mean arrival time should be equal as well. However, this expectation is not confirmed due to the disparity in the masses. In panel (a), the mass is significantly smaller than the mass in panel (b), resulting in a larger spreading factor $\omega t$ and, as a consequence, an enhancement of the temporal asymmetry induced by the term $(\omega t)^2/[1+(\omega t)^2]$. This effect can be clearly appreciated in the BM formalism through the relations \eqref{BT-free} and \eqref{pMBt-free}. As illustrated in panel (b), the asymmetry in the time-probability distribution disappears as $\omega\to 0$. This spreading behavior can also be appreciated by comparing the two panels of Fig.~\ref{f}.

To conclude, notice that the four approaches considered [${\wp_{t|q}}^{\scriptscriptstyle \mkern-25mu\tx{C}\mkern+13mu}(t|0)$, ${\wp_{t|q}}^{\scriptscriptstyle \mkern-25mu\tx{BM}\mkern+1mu}(t|0)$, ${\wp_{t|q}}^{\scriptscriptstyle \mkern-25mu\tx{L}}\,\,(t|0)$, and ${\wp_{t|q}}^{\scriptscriptstyle \mkern-25mu\tx{K}\mkern+13mu}(t|0)$] agree with each other. This test already enables us to corroborate the correctness of our formulas, at least for this simple application. Even so, we also performed the counting method described at the beginning of this section and found the same results. However, they are not shown in Fig.~\ref{f1} for a clearness reason.

%%%%%%%%%%%%%%%%%%%%%%%%%%%%%%%%%%%%%%%%%%%%%%%%%%%%%%%%%%
\subsection{Free fall under a uniform gravitational field}

Let's delve into a scenario involving a freely falling particle, where the trajectory intersects the vertical coordinate $q=0$ at two distinct instances. As discussed previously, the TET posits that the probability distribution for arrival time is governed by Eq.~\eqref{mean_2} or, upon marginalization over $p$, Eq.~\eqref{mean_class_3}. In the present subsection, however, our study will be restricted to trajectories generated by a velocity field, implying that the latter equation does not apply. In addition, we will set TET to be BM. With this choice, we remind that the Leavens result~\eqref{densidade_corrente} is retrieved~\cite{Leavens}, with the caveat that it was not derived to deal with systems involving recurrences (the backflow problem). Here, the idea is to show the differences between the two methods in characterizing the arrival-time probability distribution when this behavior manifests. Given our purposes, the analysis of the free fall motion will be effectively restricted to one dimension, where we can directly apply our formalism.

%%%%%%%%%%%%%%%%%%%
\begin{figure}[tb]
\centerline{
\includegraphics[width=11cm,angle=0]{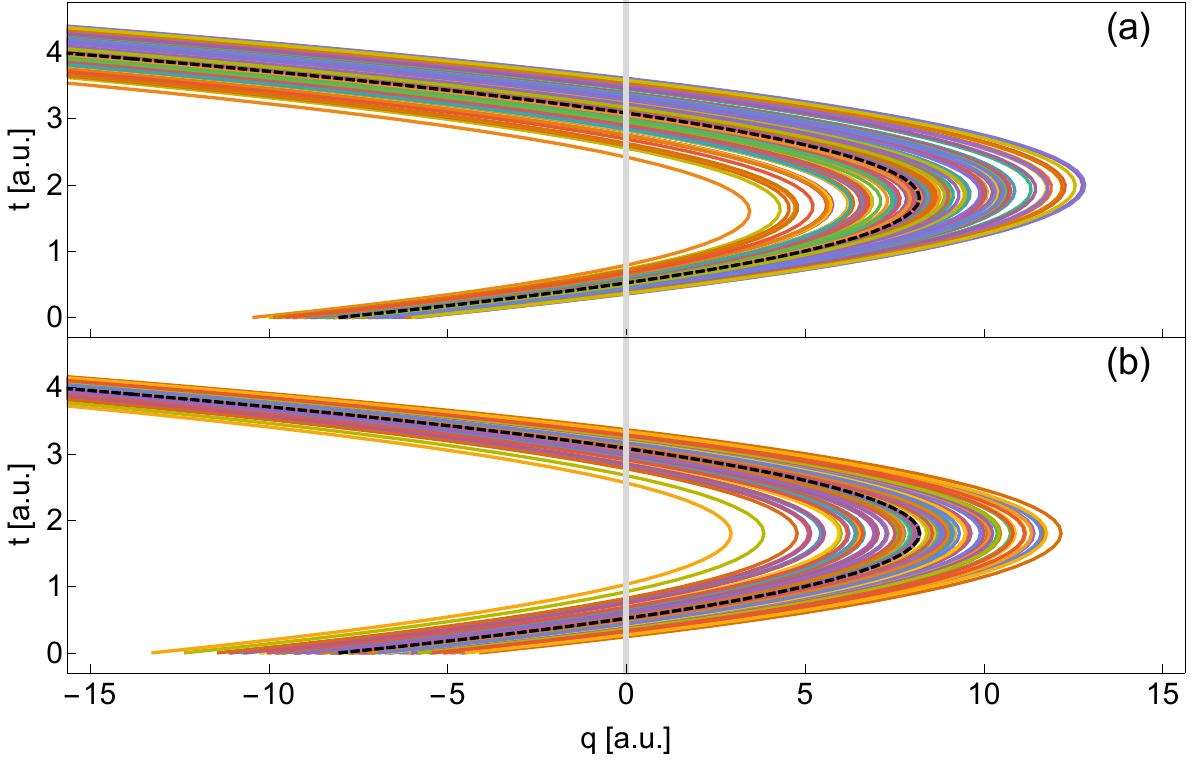}}
\caption{Bohmian trajectories of a freely falling particle for an ensemble of $N_t=100$ initial positions $q_0$. Analogously to Fig.~\ref{f}, the number of initial conditions located inside the interval $[q_0,q_0+dq_0]$ is proportional to $|\psi(q_0,0)|^2dq_0$, to be statistically representative. The black dashed curve represents the trajectory with ${q_0=\bar{q}_0}$. Panels (a) and (b) depict the distributions for two sets of parameters, namely, ${\{\sigma_0=1, \bar{q}_0=-8,\bar{p}_0=9,m=0.5,g=10,\hbar=1 \} }$ and ${ \{ \sigma_0=2,\bar{q}_0=-8,\bar{p}_0 =36,m=2,g=10,\hbar=1 \} }$, respectively. The gray line $q=0$ indicates the position where the detector is located. All physical quantities are given in arbitrary units.}
\label{f2}
\end{figure}
%%%%%%%%%%%%%%%%%%%

Consider a particle of mass $m$ immersed in a uniform gravitational field and prepared in a Gaussian state expressed by $\psi(q,t=0) =  \sqrt{G_q (\bar{q}_0, \sigma_0)}\, \mbox{e}^{i\bar{p}_0 (q-\bar{q}_0)/\hbar}$, where $\bar{q}_0$ ($\bar{p}_0$) is the mean position (momentum) and $\sigma_0$ is the initial spatial uncertainty. The Bohmian trajectory for this case is directly calculated as
\eq{\label{traj_bohmiana_queda_livre}
Q(q_0,t) = \bar{q}_0 + \frac{\bar{p}_0t}{m} -\frac{g t^2}{2} + \sqrt{1+(\omega t)^2}\,\left( q_0 - \bar{q}_0 \right),  
}
where $\omega$ is given in Eq.~\eqref{sigma_omega} and $g$ is the gravitational constant. In what follows, the parameters will be chosen to ensure that the particle crosses the coordinate $q=0$ at two distinct instants for practically all significant initial conditions $q_0$. To illustrate this behavior, in Fig.~\ref{f2}, an ensemble of Bohmian trajectories~\eqref{traj_bohmiana_queda_livre} is presented for two distinct sets of parameters, analogously to Fig.~\ref{f}. As demanded, all trajectories shown in the graphs have two passages through $q=0$.

The arrival-time probability distribution for the Leavens formalism derives from the probability current density~\cite{Muga_Leavens}. Direct calculations yield
\eq{\label{pMLt-freefall}
{\wp_{t|q}}^{\scriptscriptstyle \mkern-25mu\tx{L}\mkern+1mu}\,\,(t|q)
= \frac{1}{\mathcal{N}} |v(q,t)| \,|\psi(q,t)|^2,
}
where the normalization constant $\mathcal{N}$ is given by
\eq{\nonumber
\mathcal{N} = \int dt \, |v(q,t)| \,|\psi(q,t)|^2
}
and the velocity field takes the form
\eq{\label{fpvf}
	v(q,t)=\frac{\bar{p}_0}{m}-gt+\frac{ (\omega t)^2 }{1+(\omega t)^2}\left( \frac{q-\bar{q}_0}{t}-\frac{\bar{p}_0}{m}+\frac{gt}{2}\right),
}
with $\psi(q,t)$ being the solution to the Schr\"odinger equation. 

To apply our method~\eqref{mean_2}, we first need to invert the function~\eqref{traj_bohmiana_queda_livre} to find solutions $t_{j}=T_{\mkern-4mu j}(q,q_0)$, with $j\in\{1,2,3,4\}$. This entails solving a quartic equation for time and retaining only the two physical roots corresponding to the instants at which the trajectory crosses $q=0$ (see Fig.~\ref{f2}). To discern the physical solutions, we require that $t_j=T_{\mkern-4mu j}(q,q_0)\geq 0$, for any $\{q,q_0\}$, and the consistency condition $t=T_{\mkern-4mu j}\big(q,Q_0(q,t)\big)$. Put simply, utilizing the inversion of the initial position $q_0=Q_0(q,t)$, which is unique given $q$ and $t$, allows for finding $q_0$. If this initial condition is applied to the function $T_{\mkern-4mu j}$ and yields $t_j=t$, the solution is deemed physical for a given time instant $t$. This methodology aids in identifying the physical solutions for each time interval. Although these criteria suffice for finding the physical solutions in the current scenario, in more intricate cases, additional criteria may be necessary. For practical purposes, numerical calculations will be conducted to obtain the inverse trajectories and their derivatives, implementing them in Eq.~\eqref{mean_2}. It is worth mentioning that, for the specific case under study, the analytical expressions obtained for $t_j=T_j(0,q_0)$ do not always yield real-valued solutions. Therefore, it was necessary to appropriately choose the domain of the parameters.

%%%%%%%%%%%%%%%%%%%
\begin{figure}[!t]
\centerline{
\includegraphics[width=11cm,angle=0]{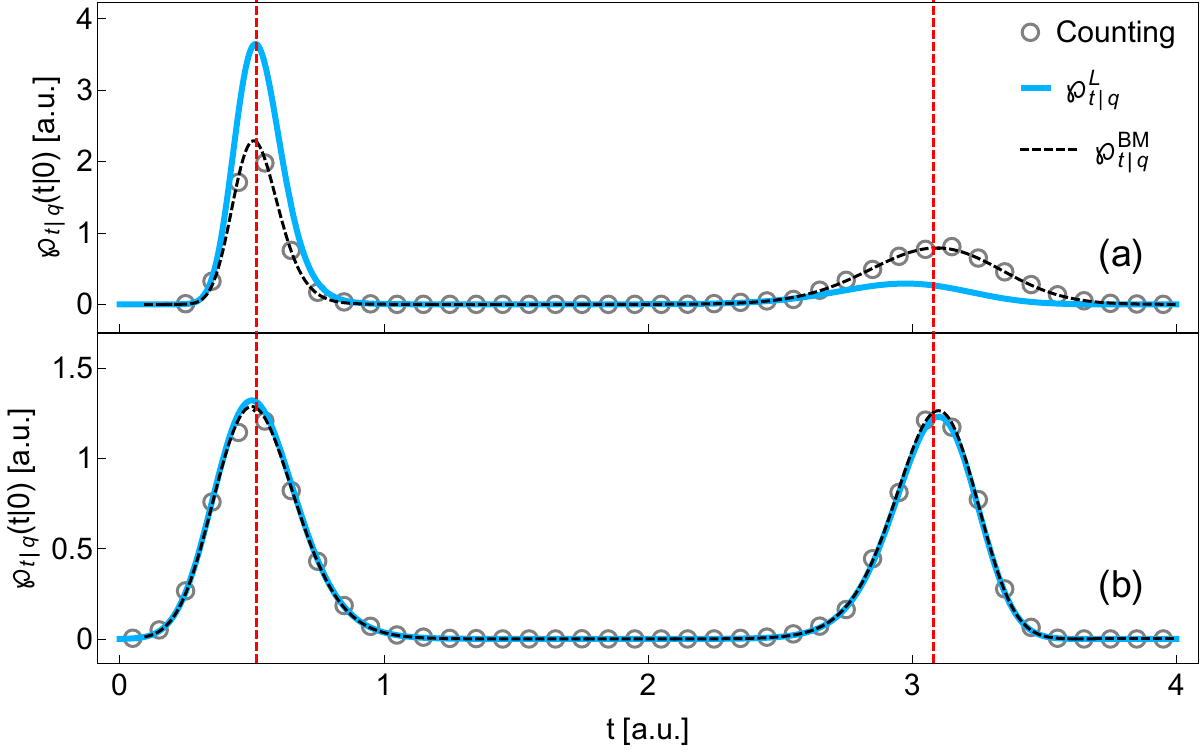}}
\caption{Arrival-time probability distributions for the problem of a freely falling particle, obtained through our formalism equipped with Bohmian trajectories [${\wp_{t|q}}^{\scriptscriptstyle \mkern-25mu\tx{BM}\mkern+1mu}(t|0)$, black dashed line], the Leavens method [${\wp_{t|q}}^{\scriptscriptstyle \mkern-25mu\tx{L}}\,\,(t|0)$, cyan solid line], and via direct counting of the number of Bohmian trajectories (gray circles). For the latter, we time evolved $5\times10^4$ initial conditions distributed according to $|\psi(q_0,0)|^2$, counting those arriving at $q=0$ after a time lapse $t$. The red dashed vertical lines show the positions of the most probable classical times ($t_1$ and $t_2$). Panels (a) and (b) display distributions for two sets of parameters, namely, ${\{\sigma_0=1, \bar{q}_0=-8,\bar{p}_0=9,m=0.5,g=10,\hbar=1 \} }$ and ${ \{ \sigma_0=2,\bar{q}_0=-8,\bar{p}_0 =36,m=2,g=10,\hbar=1 \} }$, respectively. All physical quantities are in arbitrary units.}
\label{f3}
\end{figure}
%%%%%%%%%%%%%%%%%%%

\begin{figure}[!t]
\centerline{
\includegraphics[width=14cm,angle=0]{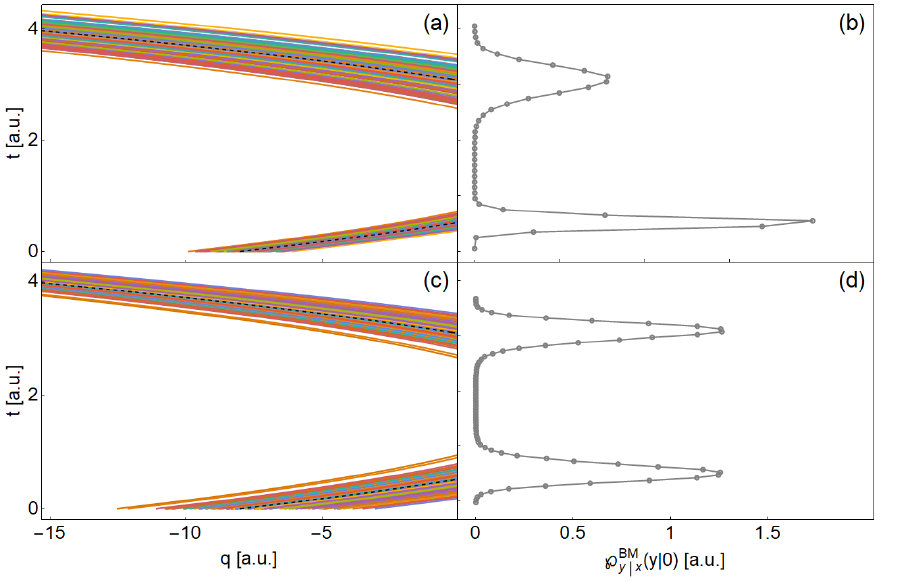}}
\caption{Bohmian trajectories copied from Fig.~\ref{f2} [panels (a) and (c)]. In panels (b) and (d), we present the results (gray circles) of the counting method using the ensembles of panels (a) and (c), respectively, offering a clear representation of such a simulation process. The lines connecting the gray circles were included to guide the eyes.}
\label{f_traj_dist}
\end{figure}

Fig.~\ref{f3} depicts the arrival-time probability distribution using the Leavens method (cyan solid curve) and our formalism equipped with Bohmian trajectories (black dashed curve). These results are also compared with the counting method explained at the beginning of this section (gray circles), where we consider an ensemble of $N_t=5\times10^4$ Bohmian trajectories. Figure~\ref{f_traj_dist} illustrates such a simulation process. We cut out the graphs of Fig.~\ref{f2}(a) and~\ref{f2}(b) at $q = 0$, showing the result in Fig.~\ref{f_traj_dist}(a) and~\ref{f_traj_dist}(c), respectively. The counts of trajectories reaching $q=0$ as a function of the arrival time give rise to the probability density functions of panels (b) and (d). Notice that the greater the density of trajectories, the higher the peaks of ${\wp_{t|q}}^{\scriptscriptstyle \mkern-25mu\tx{BM}\mkern+1mu}(t|0)$. Returning to Fig.~\ref{f3}, in both panels, the agreement between our formula~\eqref{mean_2} and the counting method implies that our analytical expressions work correctly.  Yet, the disagreement observed in panel (a) between ${\wp_{t|q}}^{\scriptscriptstyle \mkern-25mu\tx{BM}\mkern+1mu}(t|0)$ and ${\wp_{t|q}}^{\scriptscriptstyle \mkern-25mu\tx{L}}\,\,(t|0)$ is also expected, as the latter is not appropriated to describe systems exhibiting recurrences, while the former contemplates them. In this sense, we could say that ${\wp_{t|q}}^{\scriptscriptstyle \mkern-25mu\tx{BM}\mkern+1mu}(t|0)$ is more likely to be observed. We also notice that the peaks of ${\wp_{t|q}}^{\scriptscriptstyle \mkern-25mu\tx{BM}\mkern+1mu}(t|0)$ are located precisely over the most probable classical times ($t_1$ and $t_2$), that is, the instants for which the center of the Gaussian distribution reaches $q=0$. The distribution ${\wp_{t|q}}^{\scriptscriptstyle \mkern-25mu\tx{L}\mkern+1mu}\,\,(t|0)$, on the other hand, disagrees about the position of the second peak, the one associated to the recurrence time. Finally, as we approach a more classical regime (large mass), the velocity fields~\eqref{fpvf} at $t_{1,2}$ become more symmetric in the sense that  $|v(q=0,t_1)| = |v(q=0,t_2)|$. In this regime, as discussed at the end of subsection~\ref{sub:TwFoVA}, the Leavens approach becomes as effective as ours [see panel (b)]. 

%%%%%%%%%%%%%%%%%%%
%%%%%%%%%%%%%%%%%%%

The present application shows remarkable differences compared with the free-particle system. First, there is no direct application of the Kijowski formula, reducing the number of approaches to be confronted. This lack impelled us to use this model to illustrate the differences between our formula ${\wp_{t|q}}^{\scriptscriptstyle \mkern-25mu\tx{BM}\mkern+1mu}$ and ${\wp_{t|q}}^{\scriptscriptstyle \mkern-25mu\tx{L}\mkern+1mu}$, which do not coincide now due to the backflow problem. The counting approach, in this case, fulfills its role of ensuring the correctness of ${\wp_{t|q}}^{\scriptscriptstyle \mkern-25mu\tx{BM}\mkern+1mu}$. It is also important to mention that the inversion of Eq.~\eqref{traj_bohmiana_queda_livre} to get $t_{j}=T_{\mkern-4mu j}(q,q_0)$, with $j\in\{1,2,3,4\}$, is not straightforward as the analogous task of the last subsection.

%%%%%%%%%%%%%%%%%%%%%%%%%%%%%%%%%%%
\subsection{Double-slit experiment}

%%%%%%%%%%%%%%%%%%%
\begin{figure}[!t]
\centerline{
\includegraphics[width=13cm,angle=0]{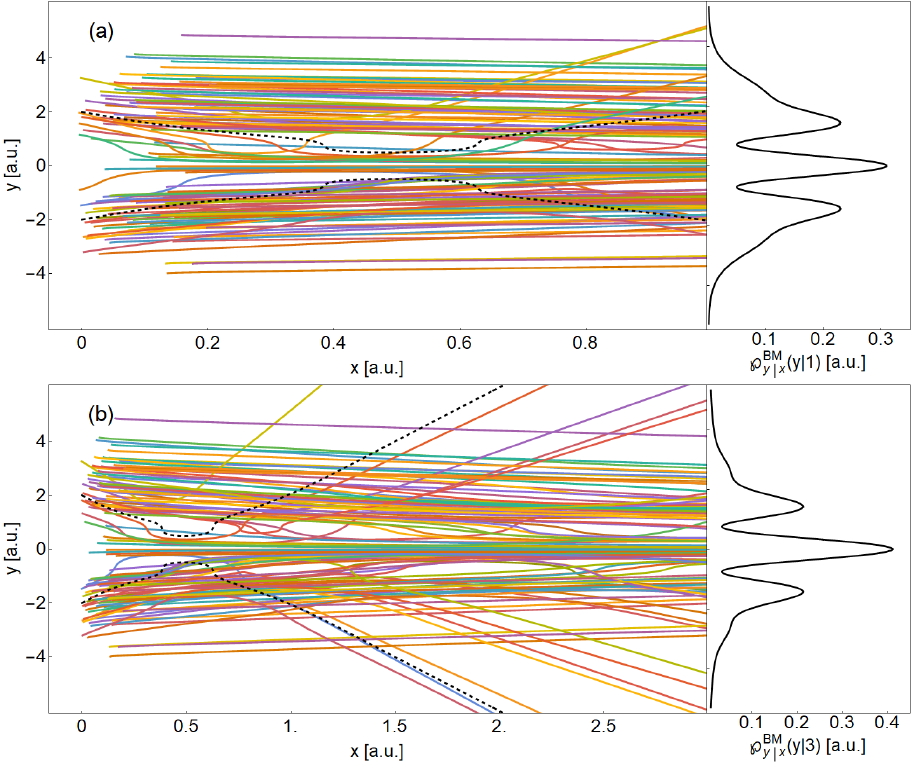}}
\caption{Left side: Bohmian trajectories for the double-slit experiment, evolved from an ensemble of $N_t=200$ initial positions $(x_0,y_0)$, according to the distribution $|\psi_x^{\scriptscriptstyle\tx{G}}(x_0,0)|^2 |\psi_y(y_0,0)|^2$. The curves observed are $y(x)$, obtained from the Bohmian trajectories, $x(x_0,t)$ and $y(y_0,t)$, with evolution time $\tau$ given by $x(x_0,\tau)= x_\tx{d}$. The trajectory starting from ${\bar{x}_0}$ and ${\bar{y}_0}$ is represented by the black dashed curve. On the right side, we present the results achieved by the numerical integration of Eq.~\eqref{teorico_dupla_fenda_1}. Panels (a) and (b) consider two sets of parameters: ${ \{ \sigma_{0x}=0.1,\sigma_{0y}=1,\bar{x}_0=0,\bar{y}_0=0,\bar{p}_{0x} =0.5,\bar{p}_{0y} =2 , m=0.5,x_\tx{d}=1, y_\tx{s}=2,\hbar=1 \} }$ and ${ \{ \sigma_{0x}=0.1,\sigma_{0y}=1,\bar{x}_0=0,\bar{y}_0=0,\bar{p}_{0x} =0.5,\bar{p}_{0y} =2 , m=0.5,x_\tx{d}=3, y_\tx{s}=2,\hbar=1 \} }$, respectively. All quantities are given in arbitrary units.}
\label{f_traj_dist_dupla}
\end{figure}
%%%%%%%%%%%%%%%%%%%

Referring back to the fundamental problem discussed in the introduction, we now apply our formalism to provide a theoretical counterpart of ${\wp_{y|x}}^{\mkern-22mu\tx{exp}}(y|x_\tx{d})$ for the double-slit experiment with particles, where the detection screen is situated in the line $(x_\tx{d},y)$. The slits occupy the positions $\pm {y}_\tx{s}$ over the line $(0,y)$. Here, we also use only BM as TET. Because the components $x$ and $y$ of the particle's motion are decoupled from each other, we have at hand two independent free-particle one-dimensional problems, which are connected only by the variable time.

For the \(x\) direction, we describe the system in two ways. First, the wave function is modeled as a Gaussian $\psi_x^{\scriptscriptstyle\tx{G}}(x,0)= \sqrt{G_x(\bar{x}_0,\sigma_{0x})}\,\mbox{e}^{i\bar{p}_{0x} \left( x-\Bar{x}_0 \right)/\hbar}$, where $G_x$ is given by Eq.~\eqref{gaussiana}. This scenario has already been studied in Sect.~\ref{sub:FP}, where we demonstrated that the TET based on Bohmian and classical trajectories, as well as the Kijowski formalism, provides the same statistical predictions for arrival times. Given this equivalence, for convenience, we adopt the TET using Bohmian trajectories to deal with this problem. Concerning the \(y\) direction, right after passing the slits, we consider
\eq{
\psi_y(y,0) = \frac{g_- \, \mbox{e}^{-i\bar{p}_{0y} (y+y_\tx{s})/\hbar} +  
g_+\,\mbox{e}^{+i\bar{p}_{0y} (y-y_\tx{s})/\hbar} }
    {\sqrt{2\left[1 + \exp\left(-\tfrac{{y}_\tx{s}^2}{2\sigma_{0y}^2}+\frac{2 \Bar{p}_{0y}^2 \sigma_{0y}^2}{\hbar^2} \right)\right]}},
    \label{psi0DSE}
}
where $g_\pm \equiv \sqrt{G_y(\pm{y}_\tx{s},\sigma_{0y})}$. Thus, it suffices to use Eqs.~\eqref{p(y|x,t)} and~\eqref{pMBt-free} to obtain the probability distribution ${\wp_{y|x}}^{\scriptscriptstyle \mkern-25mu\tx{BM}\mkern+1mu}(y|x_\text{d})\,$:
\eq{
{\wp_{y|x}}^{\scriptscriptstyle \mkern-25mu\tx{BM}\mkern+1mu}(y|x_\text{d})=
\int_0^{\infty}\!\! dt \, |\psi_y(y,t)|^2
{\wp_{t|x}}^{\scriptscriptstyle \mkern-25mu\tx{BM}\mkern+1mu}(t|x_{\text{d}}).
\label{teorico_dupla_fenda_1}
}
For added convenience, this integral is computed numerically using the Monte Carlo method. Now, to provide a Bohmian perspective on the system, we plot, in Fig.~\ref{f_traj_dist_dupla}, an ensemble of $N_t=200$ Bohmian trajectories, achieved from the initial positions, $x_0$ and $p_0$, chosen according to the distributions~$|\psi_x(x_0,0)|^2$ and $|\psi_y(y_0,0)|^2$, respectively. Once we have chosen the pair $(x_0,y_0)$, we first consider the Bohmian trajectory $x(x_0,t)$, to find the evolution time $\tau$, solution of $x(x_0,\tau)= x_\tx{d}$. Then, we calculate $y(y_0,\tau)$ to plot the curves $y(x)$ seen on the left-hand side of Fig.~\ref{f_traj_dist_dupla}. Concerning the expressions for the Bohmian trajectory $x(x_0,t)$ and $y(y_0,t)$, the former was already well studied in Sect.~\ref{sub:FP} [see Eq.~\eqref{BT-free}] while the latter demands a numerical treatment. In fact, we use the Runge–Kutta method to solve the differential equation involving $y(y_0,t)$. On the right-hand side of Fig.~\ref{f_traj_dist_dupla}, panels (a) and (b), we plot ${\wp_{y|x}}^{\scriptscriptstyle \mkern-25mu\tx{BM}\mkern+1mu}$, numerically calculated directly from Eq.~\eqref{teorico_dupla_fenda_1}.

The second approach to the present problem consists of taking the usual plane wave with box normalization to represent the state in the $x$ direction:
\eq{\psi_x^{\scriptscriptstyle\tx{PW}}(x,t) = 
\begin{cases} x_\tx{d}^{-1/2} \mbox{e}^{i(\bar{p}_{0x} x - E t)/\hbar}, & \mbox{if} \; 0 \leq x \leq x_\tx{d}, \\
0, & \mbox{if} \; x < 0 \; \text{and} \; x>x_\tx{d},
\end{cases} \label{funcao_onda_dupla}
}
where $\bar{p}_{0x}$ and $E$ are the component $x$ of the momentum and the energy of the system, respectively. For part $y$ of the wave function, we will keep Eq.~\eqref{psi0DSE}. In the $x$ direction, the Bohmian trajectory is no different from the classical one: $x =  x_0 + \bar{p}_{0x} t/m$. Using the prescriptions~\eqref{mean_1} and \eqref{mean_2}, we obtain the distributions for the final position $x$ given the arrival time $t$
\eq{
{\wp_{x|t}}^{\scriptscriptstyle \mkern-25mu\tx{BM}\mkern+1mu}(x|t)= \begin{cases}
x_\tx{d}^{-1}, & \mbox{if} \; 0 \leq x \leq x_\tx{d}, \\
0, & \mbox{if} \; x < 0 \; \text{and} \; x>x_\tx{d},
\end{cases} 
}
and the distribution for the arrival time $t$ given the final position $x$
\eq{
{\wp_{t|x}}^{\scriptscriptstyle \mkern-25mu\tx{BM}\mkern+1mu}(t|x_\tx{d})=
\begin{cases}
t_\tx{d}^{-1}, & \mbox{if} \; 0 \leq t \leq t_\tx{d}, \\
0, & \mbox{if} \; t < 0 \; \text{and} \; t>t_\tx{d},
\end{cases}
\label{tempo_dupla_fenda}
}
where $t_\tx{d}\equiv m x_\tx{d}/|\bar{p}_{0x}|$ is the classical time taken by a particle of mass $m$ that crosses the slit and moves straight with momentum $\bar{p}_{0x}$ to reach the screen. Notice that, since the spatial distribution is uniform in the interval $x\in [0,x_\tx{d}]$, the arrival-time distribution is also uniform in the time interval $t\in \left[0,t_\tx{d} \right]$. Thus, by substituting the formulas above into the prescription~\eqref{DP_yx}, we find the probability distribution of finding the particle at the final position $y$ given the final position $x_\tx{d}$
\eq{
{\wp_{y|x}}^{\scriptscriptstyle \mkern-30mu\tx{BM}\mkern+6mu}(y|x_\tx{d})=
\frac{1}{t_\tx{d}} \int_0^{t_\tx{d}}\!\! dt \, |\psi_y(y,t)|^2.
\label{teorico_dupla_fenda}
}
Here, $\psi_y(y,t)$ is calculated from the Schr\"odinger equation, and, for added convenience, the integral was computed numerically using the Monte Carlo method.

%%%%%%%%%%%%%%%%%%
\begin{figure}[!t]
\centerline{
\includegraphics[width=11cm,angle=0]{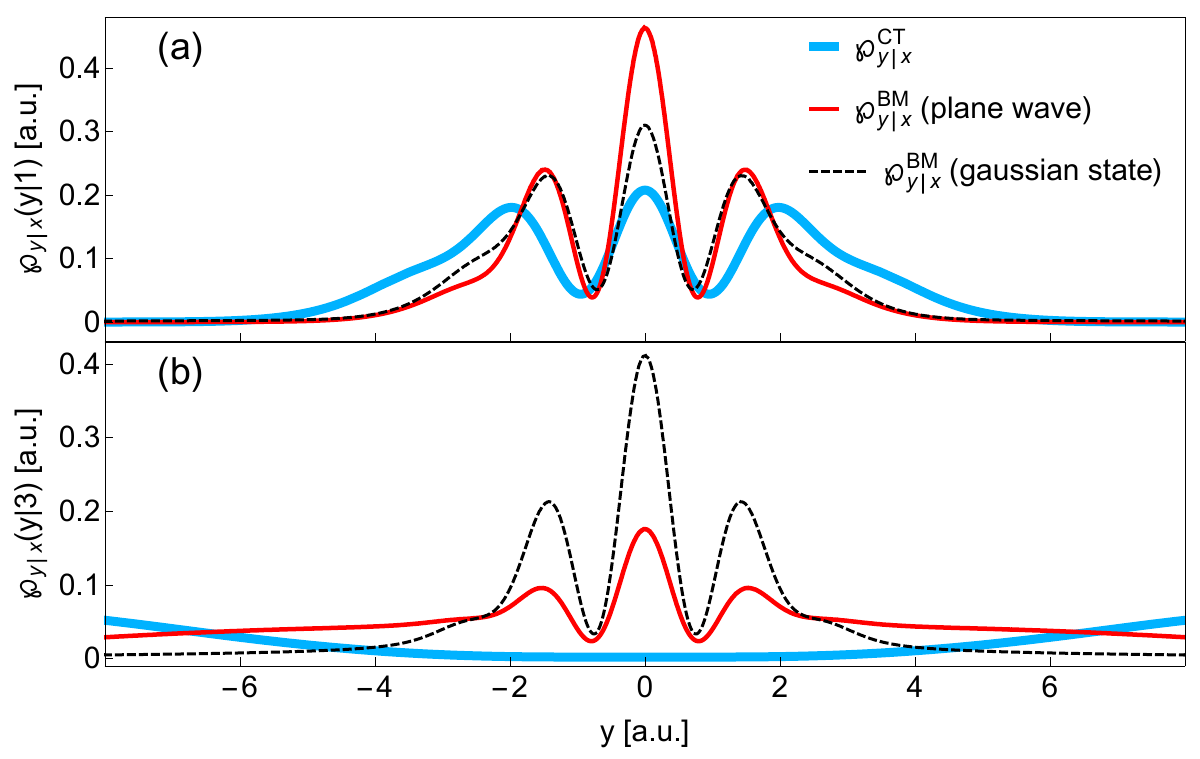}}
\caption{Probability distributions of type $\wp_{y|x}$ as a function of~$y$, for the double-slit experiment, conditioned to a detection at the coordinate $x=x_\tx{d}$. The black dashed curve represents the results from our formalism ${\wp_{y|x}}^{\scriptscriptstyle \mkern-27mu\tx{BM}\mkern+3mu}(y|x_\tx{d} )$ using Gaussian wavepacket in the direction $x$, the red solid line ${\wp_{y|x}}^{\scriptscriptstyle \mkern-27mu\tx{BM}\mkern+3mu}(y|x_\tx{d} )$ was built using the plane wave, and the cyan solid curve represents the results from quantum mechanics supplemented with classical time ${\wp_{y|x}}^{\scriptscriptstyle \mkern-27mu\tx{CT}\mkern+3mu}(y|x_\tx{d} )$. Panels (a) and (b) depict the distributions for two sets of parameters: ${ \{ \sigma_{0x}=0.1,\sigma_{0y}=1,\bar{x}_0=0,\bar{y}_0=0,\bar{p}_{0x} =0.5,\bar{p}_{0y} =2 , m=0.5,x_\tx{d}=1, y_\tx{s}=2,\hbar=1 \} }$ and ${ \{ \sigma_{0x}=0.1,\sigma_{0y}=1,\bar{x}_0=0,\bar{y}_0=0,\bar{p}_{0x} =0.5,\bar{p}_{0y} =2 , m=0.5,x_\tx{d}=3, y_\tx{s}=2,\hbar=1 \} }$, respectively, where the only difference is the distance $x_\tx{d}$ from the detection screen to the slits. All physical quantities are in arbitrary units.}
\label{f4}
\end{figure}
%%%%%%%%%%%%%%%%

In search of a counterpart to our result, we now supplement QM with a (semi)classical element recognized as suitable in far-field regions for certain experiments \cite{Vona,Das,Shucker,Wolf}. Specifically, we just substitute the classical arrival time, $t_\tx{d}=m x_\tx{d}/|\bar{p}_{0x}|\equiv t_\tx{d}(x_\tx{d})$, in the prescription \eqref{p(y|x,t)}, in order to obtain ${\wp_{y|xt}}\big(y|x_\tx{d},t_\tx{d}(x_\tx{d})\big)$, here denoted as
\eq{\label{pCTyx}
{\wp_{y|x}}^{\scriptscriptstyle \mkern-30mu\tx{CT}\mkern+6mu}(y|x_\tx{d} )
= | \psi_y\big(y,t_\tx{d}(x_\tx{d})\big)|^2,
}
with ``CT'' standing for classical time. Notice that $t_{\tx{d}}$ has the same value for both treatments $\psi_x^{\scriptscriptstyle\tx{G}}$ and $\psi_x^{\scriptscriptstyle\tx{PW}}$, implying the equivalent conclusion for the classical distribution ${\wp_{y|x}}^{\scriptscriptstyle \mkern-30mu\tx{CT}\mkern+6mu}(y|x_\tx{d} )$. 

In Fig.~\ref{f4}, the results obtained with our formalism, using BM as TET, are plotted as the black dashed (Gaussian case, $\psi_x^{\scriptscriptstyle\tx{G}}$) and red solid (plane-wave case, $\psi_x^{\scriptscriptstyle\tx{PW}}$) curves, while the quantum-mechanical approach supplemented with a classical arrival time is plotted as a cyan solid line. Although not always accentuated, the differences between the models ${\wp_{y|x}}^{\scriptscriptstyle \mkern-30mu\tx{BM}\mkern+6mu}(y|x_\tx{d} )$ and ${\wp_{y|x}}^{\scriptscriptstyle \mkern-30mu\tx{CT}\mkern+6mu}(y|x_\tx{d} )$ are perceptible and, in principle, measurable. In panel (a), where the two Gaussian branches of the wave function~\eqref{psi0DSE} overlap significantly, all models predict the central interference pattern, but with different visibilities. In panel (b), which represents a situation with low overlap, the differences are remarkable and can be traced back to formulas~\eqref{teorico_dupla_fenda_1}, \eqref{teorico_dupla_fenda} and \eqref{pCTyx}. When we take $x_\tx{d}$ to be large, and thus the arrival time $t_\tx{d}=mx_\tx{d}/|p_{0x}|$, the latter formula allows the branches of the wave function to move far apart, reducing the overlap and preventing the central interference pattern from manifesting itself. On the other hand, formulas~\eqref{teorico_dupla_fenda_1} and~\eqref{teorico_dupla_fenda} prescribe time averaging over several times, thus preserving the central interference pattern, albeit with reduced visibility. In other words, from an experimental standpoint, ${\wp_{y|x}}^{\scriptscriptstyle \mkern-30mu\tx{CT}\mkern+6mu}(y|x_\tx{d} )$ corresponds to statistically post-selecting ${\wp_{y|x}}^{\scriptscriptstyle \mkern-30mu\tx{BM}\mkern+6mu}(y|x_\tx{d} )$ for the specific time value $t_\tx{d}$. As for the counting result, which is not shown in the graphs for clearness reasons, it is noteworthy that it consistently corroborates our statistical formalism, which is based on ${\wp_{y|x}}^{\scriptscriptstyle \mkern-30mu\tx{BM}\mkern+6mu}(y|x_\tx{d} )$.

To make the above discussion more precise, we now introduce the visibility $\mathcal{V}$ of an interference pattern, which is a common way to quantify fringe contrast. In particular, we calculate the visibility of the central fringe through
\eq{\label{visibilidade}
\mathcal{V}_0 = \frac{\max_0[\wp_{y|x}(y|x_{\text{d}})] - \min_0[\wp_{y|x}(y|x_{\text{d}})]}{\max_0[\wp_{y|x}(y|x_{\text{d}})] + \min_0[\wp_{y|x}(y|x_{\text{d}})]},
}
where \(\max_0[\wp_{y|x}(y|x_{\text{d}})]\) denotes the maximum value of $\wp_{y|x}(y|x_{\text{d}})$ around $y=0$, and \(\min_0[\wp_{y|x}(y|x_{\text{d}})]\) corresponds to its minimum value at the adjacent valleys. We calculated the visibility for each probability distribution shown in Fig.~\ref{f4}. For panel (a), we obtain \(\mathcal{V}_0^{\text{BM}}\,(\text{Gaussian state}) \approx 0.70\), \(\mathcal{V}_0^{\text{BM}}\,(\text{plane wave}) \approx 0.84\), and \(\mathcal{V}_0^{\text{CT}} \approx 0.65\). For panel (b), we get \(\mathcal{V}_0^{\text{BM}}\,(\text{Gaussian state}) \approx 0.85\), \(\mathcal{V}_0^{\text{BM}}\,(\text{plane wave}) \approx 0.76\), and \(\mathcal{V}_0^{\text{CT}} \approx 0\), the latter implying that no interference fringe is observed around $y=0$. In double-slit experimental setups, such as those described in~\cite{Zeilinger, Gahler, Tonomura, Bach, Carnal, Kurtsiefer, Brand}, the intensity of the particle distribution is measured, from which the visibility can be calculated and compared with theoretical predictions. Furthermore, a direct comparison between experimental histograms and theoretical distributions is, in principle, also possible.

As a final remark, we note that the reduced visibility exhibited by ${\wp_{y|x}}^{\scriptscriptstyle \mkern-30mu\tx{BM}\mkern+6mu}(y|x_\tx{d} )$ in Fig.~\ref{f4} (their minima do not reach the horizontal axis) might lead one to believe, when confronted with experimental data of this form, that the system is subjected to some source of decoherence. However, what is being shown here is that this effect derives solely from arrival-time fluctuations; no spurious, uncontrollable degrees of freedom have been considered in our treatment. The behavior of ${\wp_{y|x}}^{\scriptscriptstyle \mkern-30mu\tx{BM}\mkern+6mu}(y|x_\tx{d} )$ for the plane-wave case, reducing the visibility from panel (a) to (b), also suggests this possibility. Indeed, as $t_\tx{d}$ and the integration range of integral~\eqref{teorico_dupla_fenda} increases in the second plot, those fluctuations should also increase, as should the visibility. This same behavior does not manifest for the Gaussian case, but it follows from the fact that integral~\eqref{teorico_dupla_fenda_1} assigns a (Gaussian) localized statistical weight to each arrival time, alleviating the effects of fluctuations. This discussion raises an intriguing question about low-visibility experimental patterns. If one could guarantee the isolation of the system, then the reduced visibility could be seen as corroborating our theory.

In this subsection, we have faced the system that is the main motivation of our work. Overcoming the technical difficulties to (numerically) integrate the Bohmian trajectories, we achieved results distinguishable from the most commonly used ${\wp_{y|x}}^{\scriptscriptstyle \mkern-30mu\tx{CT}\mkern+6mu}(y|x_\tx{d})$, which has the potential to stimulate efforts to perform experimental studies. Also, notice that the Kijowski method here is just a supporting character, agreeing with ${\wp_{x|t}}^{\scriptscriptstyle \mkern-30mu\tx{BM}\mkern+6mu}$ in the part of the problem where it applies. The same happens with Leavens formula ${\wp_{x|t}}^{\scriptscriptstyle \mkern-25mu\tx{L}\mkern+6mu}$, as the treatment for the $x$-direction has no backflow behavior.

%%%%%%%%%%%%%%%%%%%%%%%
\section{Final remarks}
\label{sec:conclusion}

In this work, we demonstrate that trajectory-equipped theories have a potential advantage over QM in predicting the statistical outcomes of experiments involving arrival-time fluctuations. Specifically, we show that by starting with the determinism hypothesis, a formalism can be developed [see Eqs.~\eqref{mean_1}, \eqref{mean_2}, \eqref{mean_class_1}, \eqref{mean_class_2}, and \eqref{mean_class_3}] that yields predictions for the arrival-time probability distributions in emblematic experiments. Most importantly, we have suggested a way to test trajectory-based determinism using a conditional probability distribution [Eq.~\eqref{DP_yx}], which involves position measurements only, as usually done in experimental setups. If confirmed empirically, our predictions will provide support for the idea of an underlying trajectory-based determinism spanning the quantum realm. On the other hand, if these predictions are not confirmed, then the trajectory-equipped theory being tested would be ruled out. We hope that this work may stimulate researchers to test our results in the laboratory, which, we believe, is feasible with current technology.

%%%%%%%%%%%%%%%%%%%%%%%%%
\section{ACKNOWLEDGMENTS}

This study was financed in part by the Coordenação de Aperfeiçoamento de Pessoal de Nível Superior - Brasil (CAPES) - Finance Code 001. R.M.A. and A.D.R. thank the financial support from the National Institute for Science and Technology of Quantum Information (CNPq, INCT-IQ 465469/2014-0). R.M.A. also thanks the Brazilian funding agency CNPq for Grant No. 305957/2023-6. A.D.R acknowledges Lea Santos and the Uconn Physics Department for their hospitality during his stay in Storrs, where part of this work was done.

%----------------------------
%----------------------------

\end{document}